\documentclass{article}

\usepackage{arxiv}

\usepackage[utf8]{inputenc} 
\usepackage[T1]{fontenc}    
\usepackage{amsmath}
\usepackage{hyperref}       
\usepackage{url}            
\usepackage{booktabs}       
\usepackage{amsfonts}       
\usepackage{nicefrac}       
\usepackage{microtype}      
\usepackage{cleveref}       
\usepackage{lipsum}         
\usepackage{graphicx}
\usepackage{natbib}
\usepackage{doi}

\title{Accuracy Enhancement for Ear Acoustic Authentication Using Between-class Features}

\author{ \href{https://orcid.org/0000-0002-8371-2868}{\includegraphics[scale=0.06]{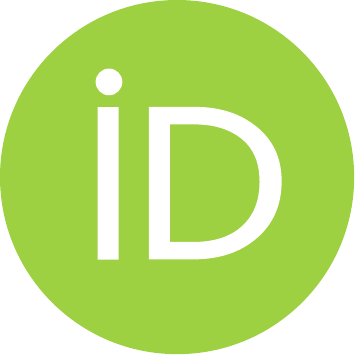}\hspace{1mm}Masaki YASUHARA} \\
Department of Science of Technology Innovation\\
Nagaoka University of Technology 1603-1,\\
Kamitomioka Nagaoka, \\
Niigata 940-2188, Japan\\
\texttt{myasuhara@stn.nagaokaut.ac.jp} \\
\And
\href{https://orcid.org/0000-0002-1705-6268}{\includegraphics[scale=0.06]{orcid.pdf}\hspace{1mm}Isao Nambu} \\
Graduate School of Engineering, \\
Nagaoka University of Technology 1603-1, \\
Kamitomioka Nagaoka, \\
Niigata 940-2188, Japan\\
\And
Yoshiko MARUYAMA \\
National Institute of Technology, \\
Hakodate College\\
14-1 Tokuracho, Hakodate,\\
Hokkaido, 042-0953,  Japan\\
\And
Shohei YANO \\
Department of Electrical and Electronic Systems Engineering,\\
National Institute of Technology, Nagaoka College \\
888 Nishikatakai Nagaoka\\
Niigata 940-8532, Japan
}

\renewcommand{\shorttitle}{Ear Acoustic Authentication Using Between-class Features}

\hypersetup{
pdftitle={Accuracy Enhancement for Ear Acoustic Authentication Using Between-class Features},
pdfsubject={q-bio.NC, q-bio.QM},
pdfauthor={Masaki Yasuhara et. al.},
pdfkeywords={Biometrics, Data Augmentation, More},
}

\begin{document}
\maketitle

\begin{abstract}
	In existing biometric authentication methods, the user must perform an authentication operation such as placing a finger in a scanner or facing a camera.
	With ear acoustic authentication, acoustic characteristics of the ear canal are used as biometric information.
	Therefore, a person wearing earphones does not need to perform any authentication operation.
	In biometric authentication, it is necessary to minimize the false acceptance rate (FAR) so that no unauthorized user is misidentified as an authorized user.
	However, if the FAR is set low, it increases the false rejection rate (FRR), the rate at which authorized users are falsely recognized as unauthorized users.
	It has been reported that when FAR is 0.1\%, the FRR in ear acoustic authentication reaches as much as 22\%.
	In this study, we propose a method that reduces FRR and enhances authentication accuracy; it generates new ear canal acoustic characteristics called between-class (BC) features, which combine the ear canal acoustic characteristics of authorized and unauthorized users features.
	The proposed method uses a support vector machine to learn the BC features as the data of authorized users, then generates a hyperplane in an area close to that data.
	We hypothesize that this would decrease the possibility of misidentifying an unauthorized user as an authorized user and decrease the FRR when the FAR is 0.1\%.
	To evaluate the performance of the proposed method, BC features were applied to ear acoustic authentication, and FAR and FRR were calculated.
	The FRR with FAR = 0.1\% was 7.95\% lower than with the previous method, and the equal error rate -- the error rate when FAR and FRR are equivalent -- decreased by 0.15\%.
	These results confirmed that the proposed method can make ear acoustic authentication more convenient while maintaining a high level of security. 
	This paper is under peer-previewing of IEICE Trans. Inf. \& Syst.
\end{abstract}


\section{Introduction}
Biometric authentication can minimize the risks of information loss, theft, or leak. Over the past few decades, many types of biometric authentication have been studied, including recognition by fingerprint, face, iris, retina, and voice \cite{Mizoguchi2010,Imaoka2010,Koshinaka2015}.
These methods require an authentication operation from the user, such as placing a finger on a scanner or facing a camera.
We focused on the ear canal as new biometric information representing human individuality.
Using the ear canal for identification enables a type of biometric authentication that does not require any authentication operation; all it requires is to wear an earphone-shaped authentication device.
In the field of sound image localization technology -- which virtually creates the direction of arrival of sound for earphones and headphones -- previous studies have indicated that individual differences in the outer shape of the ear and the shape of the ear canal manifest as differences in how people hear sounds \cite{Moller1992,onzou,Bulauert1997}.
One of the studies on methods of personal identification using the ear canal examined the possibility of identifying individuals by emitting acoustic signals towards the ears and measuring the acoustic characteristics reflected by the shape of the pinna and the ear canal \cite{Akkerman2005}.
However, there were issues with the accuracy of the measurement of acoustic characteristics and extraction of feature amount that could be effectively used for authentication.
As a result, the expected level of authentication accuracy was not reached, and this method has not yet been put into practice. 

Using an impulse response measurement method that uses pseudo-noise signals \cite{Borish1983}, we previously demonstrated a method of personal authentication that captures sounds reflected in the ear canal with high accuracy and uses them as features for personal identification and confirmed the effectiveness of this method through an experiment that employed subjects \cite{Yano2015}.
The type of biometric authentication that uses this method \cite{Yano2015}, is called ear acoustic authentication in this paper. 
Various studies have sought ways to enhance the accuracy of biometric authentication using acoustic characteristics of the ear canal. 
One of these studies demonstrated that by minimizing measurement errors due to background noise and producing a constant signal-to-noise ratio (SNR) across all frequency bands, it is possible to capture ear acoustic authentication features very accurately \cite{yano2017}.
Other studies showed the effectiveness of a method that reduces observation fluctuations by additive averaging \cite{Kaneda2008,yano2017}.
A method of ear acoustic authentication using inaudible sounds was examined by Mahto et al., who built a hybrid system that uses both audible and inaudible areas \cite{shivangi2018}.
These ear acoustic authentication methods used the features of only one ear; we demonstrated that combining features of both ears increases authentication accuracy \cite{yasuhara2019}.

Misidentification of an unauthorized user as an authorized user is a major issue in biometrics authentication.
In ear acoustic authentication, when the false acceptance rate (FAR)-- the probability of incorrectly accepting unauthorized users -- is 0.1\%, the false rejected rate (FRR) -- the probability of incorrectly rejecting an authorized user -- is 22\% \cite{yasuhara2019}.
That is, if a biometric authentication system is operated at 0.1\% FAR, it incorrectly recognizes an authorized user as an unauthorized user once every five times.
Therefore, it is necessary to enhance the accuracy of ear acoustic authentication to make it more convenient while maintaining a high security level.
In this study, we propose a method that improves authentication accuracy by generating between-class (BC) features that combine the ear acoustic features of users.
In this study, we generate BC features for a dataset of 52 subjects, carry out 1:1 (one-to-one) authentication using a support vector machine (SVM) and empirically evaluate its effectiveness. 

\section{Ear Acoustic Authentication System}
\begin{figure}[tb]
	\centering
	\includegraphics[width=0.5\linewidth]{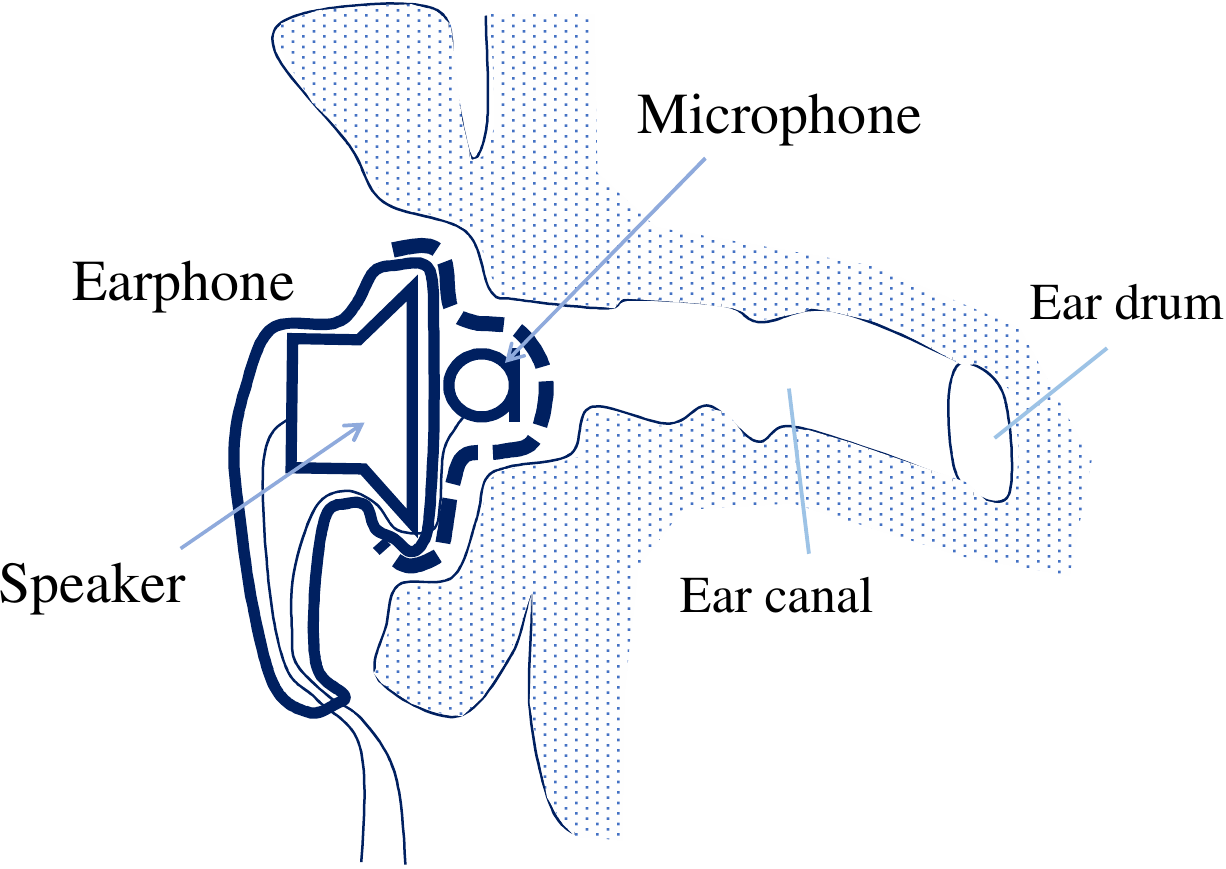}
	\caption{A special earphone which has a speaker and a microphone}
	\label{fig:eaasystem}
\end{figure}

In ear acoustic authentication, biometric authentication is performed using a special earphone with a small microphone embedded, as illustrated in Fig. \ref{fig:eaasystem}.
It emits a maximum-length sequence (MLS) signal from the speaker inside the earphone and the microphone records the sound echoed from the ear canal to capture its acoustic characteristics. 
These ear canal acoustic characteristics include reflection, diffraction, interference, and resonance characteristics, which vary from person to person according to the shape of the ear canal and the body.
To perform biometric authentication, ear acoustic authentication focuses on the individuality of these ear canal acoustic characteristics \cite{Wenzel1993,Yano2000}, which go through feature amount extraction and are applied to a discriminator. 

\begin{figure*}[tb]
	\centering
	\includegraphics[width=0.9\linewidth]{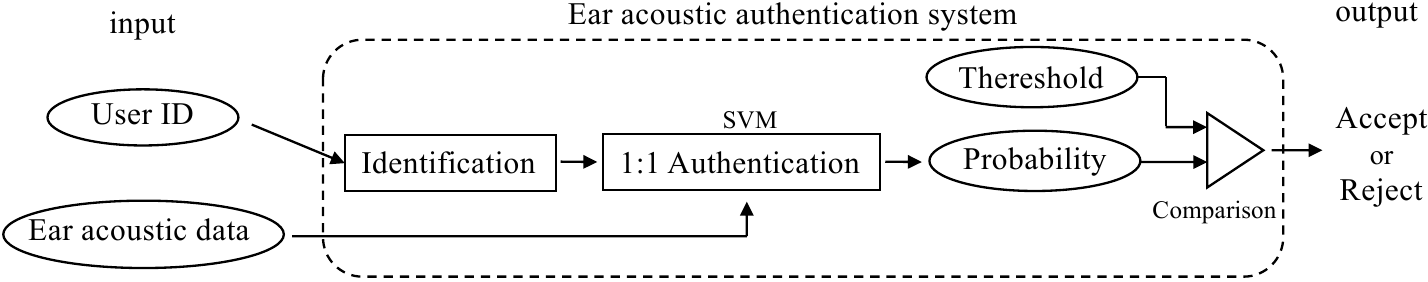}
	\caption{Overview of ear acoustic authentication system}
	\label{fig:oveaas}
\end{figure*}

In this study, we assume the ear acoustic authentication system illustrated in Fig. \ref{fig:oveaas}.
This system runs with 1:1 authentication, a type of authentication that initially identifies the user, then identifies whether the measured biometric information belongs to that user.
As the discriminator of authentication, we use a support vector machine (SVM), a type of supervised learning used for classification, regression, and outlier detection.
It can also calculate posterior probabilities as suggested by Platt al. \cite{Platt99probabilisticoutputs} and  Wu al. \cite{Wu2004}.
If the posterior probability is high, the probability that it is the identified user is high.
Ultimately, it compares the posterior probability to a threshold and outputs “Accept” or “Reject”. 

\subsection{Factors of Evaluation}
\begin{figure}[tb]
	\centering
	\includegraphics[width=0.8\linewidth]{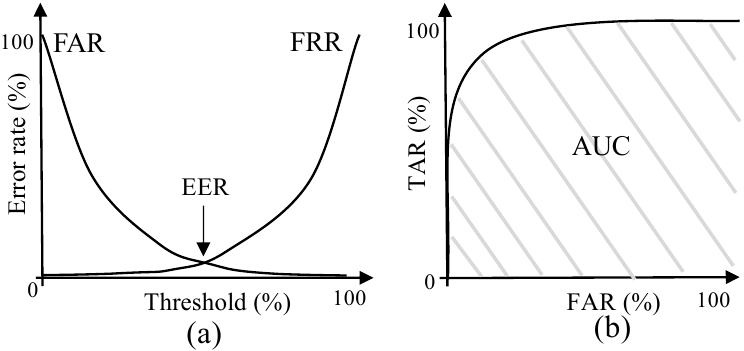}
	\caption{DET curve and ROC curve. (a) DET curve : FAR and FRR against threshold. (b) ROC curve : TAR against FAR.}
	\label{fig:det}
\end{figure}

In 1:1 authentication, the biometric information provided is classified as belonging to an authorized user or not. 
As mentioned before, FAR is the rate at which biometric information belonging to an unauthorized user is incorrectly accepted as that of an authorized user, and FRR is the rate at which biometric information belonging to an authorized user is incorrectly rejected as that of an unauthorized user.
True acceptance rate (TAR) is the rate at which biometric information of an authorized user is correctly accepted as that of an authorized user. These are defined by the following equations: 
\begin{align}
	FAR &= \frac{T_{FA}}{N_{auth}}\\
	FRR &= \frac{T_{FR}}{N_{auth}}\\
	TAR &= 1 - FRR
\end{align}
Here, $N_{auth}$ is the number of authentications performed, $T_{FR}$ is the number of authorized users incorrectly rejected, and $T_{FA}$ is the number of unauthorized users incorrectly accepted.
When doing the classification, it is necessary to set a threshold that determines the minimum probability of being an authorized user to be accepted.
If this threshold is set low, it is more difficult to incorrectly reject an authorized user and easier to incorrectly accept an unauthorized user.
On the contrary, if the threshold is set high, it becomes easier to reject authorized users.
Therefore, as indicated in Fig. \ref{fig:det}(a), there is a trade-off between FAR and FRR.
The graph shown in Fig. \ref{fig:det}(a), indicating the relationship between FAR and FRR, is known as a detection error tradeoff (DET) curve. 
The intersection between FAR and FRR is called the equal error rate (EER), which is used as a performance indicator of biometric authentication.
The lower the EER, the higher the authentication accuracy.
The graph shown in Fig. \ref{fig:det}(b), indicating the relationship between TAR and FAR, is known as a receiver operating characteristic (ROC) curve.
The area below the ROC curve is called area under the curve (AUC) and it is also used as a performance indicator of biometric authentication.
AUC varies from 0 to 1; if there were no authentication mistakes, AUC is 1, and in the case of accidental authentication, 0.5. 
Therefore, the higher the AUC, the higher the authentication accuracy. 
%

\section{Between-class feature and related works}
BC features is the feature amount combining the training samples of authorized users, $x_1$, and training samples of unauthorized users, $x_2$, with ratio $r(0<r<1)$ as per the following equation:  
\begin{equation}
	\label{eq:mixr}
	r x_1 + (1-r) x_2
\end{equation}
Here, $\bf{X_1}$ is the group of training samples of authorized users, $\bf{X_2}$ is the group of training samples of unauthorized users, so $x_1 \in \bf{X_1}$ and $x_2 \in \bf{X_2}$.
We considered the possibility of learning the BC features as unauthorized users when performing 1:1 authentication by SVM and thereby reducing the FAR of ear acoustic authentication.
The expected effect of performing 1:1 authentication on two-dimensional data of authorized and unauthorized users using an SVM can be explained as follows.
In previous ear acoustic authentication, a hyperplane that maximizes the margin of each feature amount, in the position indicated by the dotted line in Fig. 4, was chosen.
However, when generating BC features and treating them as unauthorized users, the hyperplane is expected to move closer to the data of authorized users.
With this, it becomes more difficult to incorrectly accept unauthorized users, and if the threshold of the biometric authentication system is set low, fewer unauthorized users would be incorrectly accepted. 

Therefore, the FAR curve is expected to drop rapidly as the threshold increases, as illustrated in Fig. \ref{fig:bc-det}.
On the other hand, when the hyperplane gets closer to the data of authorized users, the probability of incorrectly rejecting authorized users could increase.
In other words, the steepness of the entire FRR curve is expected to increase.
However, how the FAR and FRR curves change because of BC features is unknown.
The ratio $r$ and $N_{BC}$ (the number of BC features) are parameters thought to be closely related to authentication accuracy.
Since BC features generate features in the middle of two classes, ideally, the difference between the features ($x_1$ and$x_2$) that will generate the BC features should be small.
Therefore, the proposed method calculates the cosine similarity between $x_1$ and $x_2$ and generates $N_{BC}$ units of BC features from the combination with high cosine similarity, as in Eq. \ref{eq:cossim}:
\begin{equation}
	\label{eq:cossim}
	\cos(x_1, x_2) = \frac{x_1 \cdot x_2}{|x_1||x_2|}
\end{equation}
\begin{figure}[tb]
	\centering
	\includegraphics[width=0.7\linewidth]{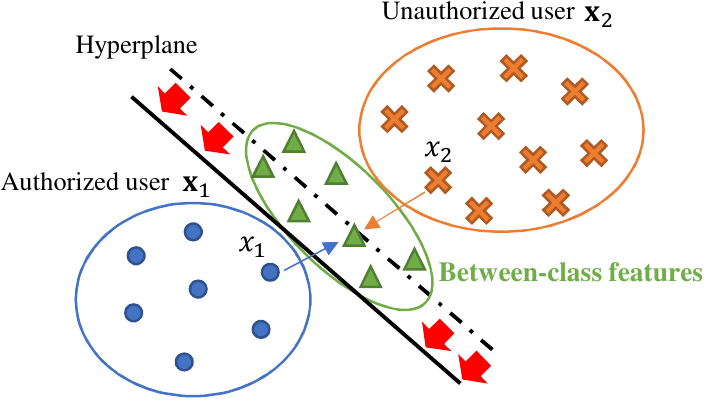}
	\caption{Hyperplane in feature space that changes with the introduction of BC features}
	\label{fig:bc-fs}
\end{figure}
\begin{figure}[tb]
	\centering
	\includegraphics[width=0.6\linewidth]{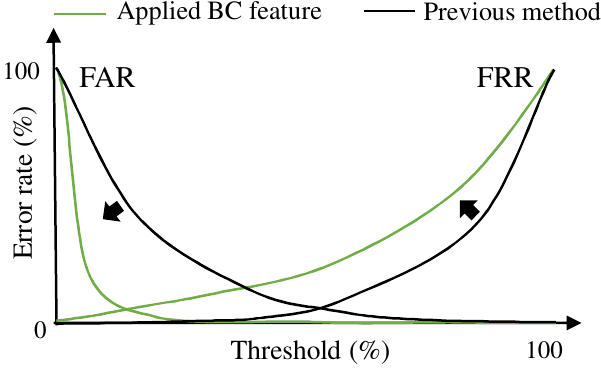}
	\caption{The lower FAR and higher FRR with introducing of BC features  compared to previous method which single ear data without BC}
	\label{fig:bc-det}
\end{figure}

Other studies have sought to improve the classification accuracy by combining two training samples.
Mix-up \cite{Zhang2017} is a data augmentation method that mixes pairs of two training samples to generate new training samples.
Between-class learning \cite{tokozume18} is a data augmentation method that mixes training sample pairs of two different classes to generate new training samples.
The ratio $r$ used to combine two feature amounts of mix-up is a random number that follows a beta distribution, and in between-class learning, it is a random number that follows a uniform distribution. 
Mix-up and between-class learning are data augmentation methods in deep learning.
In these methods, training samples combining two features with random proportions are treated as new classes.
This smooths the decision boundaries and makes over-learning difficult to occur.
Meanwhile, the proposed method aims to bring the separation interface in 1:1 authentication closer to the data of authorized users.
By learning training samples that combine two features with a constant rate as unauthorized users, this method brings the separation interface closer to the data of authorized users to reduce the FAR.
Therefore, the related studies and the proposed method have different objectives, particularly the fact that the ratio $r$ used to combine two features is a constant in the proposed method.

\section{Experimental methods}
To evaluate the effect of introducing BC features, 1:1 authentication was performed using the dataset of ear acoustic authentication used in a previous study \cite{yasuhara2019}. 
It was then evaluated by calculating the FAR, FRR, EER, and AUC. 

\subsection{Dataset}
\begin{figure*}[tb]
	\centering
	\includegraphics[width=0.75\paperwidth]{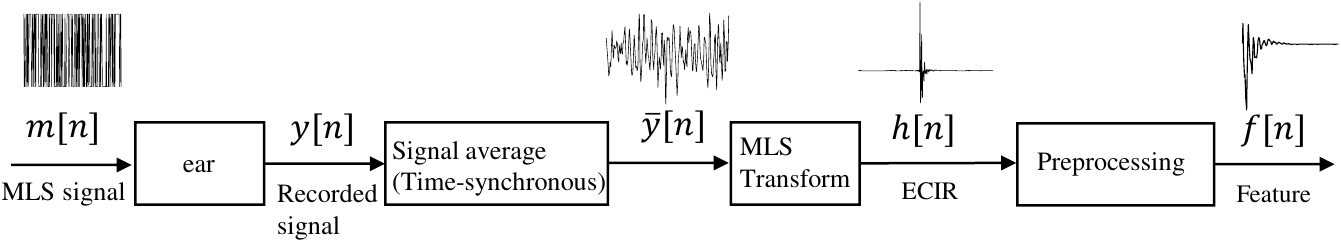}
	\caption{The flow of data measurement and preprocessing}
	\label{fig:sokuteikei}
\end{figure*}

The data used in this experiment are the same data used in a previous study \cite{yasuhara2019}.
We enrolled 52 students and employees of the National Institute of Technology, Nagaoka College aged between late teens and 50s.
Following the college rules, we obtained informed consent from all subjects to participate in this study.
The flow of the measurement of ear canal acoustic characteristics and preprocessing is shown in Fig. \ref{fig:sokuteikei}.
The ear canal acoustic characteristics of each person were measured 30 times using a canal-type earphone (Bose SoundTrue Ultra).
An MLS signal $m[n]$ (signal length $n = 2^{14}-1$) was emitted from the earphone with a noise level of 65 dB(A) and its reflected sound $y[n]$ was measured with a small microphone.
This measurement was done five times and the averaged signal $\bar{y}[n]$ was subject to Hadamard transform to derive the impulse response of acoustic transmission characteristics $h[n]$ of the ear canal.
The feature amount of both ears was captured, but since only one ear was enough to check the effectiveness of BC features, we used the data from the left ear of all subjects. 

\subsection{Preprocessing}
When minimum-phase transformation using Hilbert transform is applied to $h[n]$, the impulse response components comprising the features were found in the head of the signal.
After normalizing the power of all signals to 1 and clipping 256 samples from the head, bandpass filtering (100 Hz–22 kHz) was performed. This is used as the feature amount $f[n]$.

\subsection{Classification}
As mentioned before, this study is based on 1:1 authentication, which identifies whether the feature amount input belongs to an authorized user or an unauthorized user.
The classification is performed using the SVC function, an SVM for classification implemented in scikit-learn \cite{scikit-learn}.
The SVC function was implemented based on libsvm \cite{CC01a} and can specify the kernel and regularization parameters used.
The data of authorized users were divided into six samples of training data and 24 samples of test data.
For the data of unauthorized users, we excluded the data of one person for test data, and the rest data for learning.
%
BC features were generated from the training data of authorized users and unauthorized users and were treated as training data of unauthorized users. 

\begin{figure}[tb]
	\centering
	\includegraphics[width=0.6\linewidth]{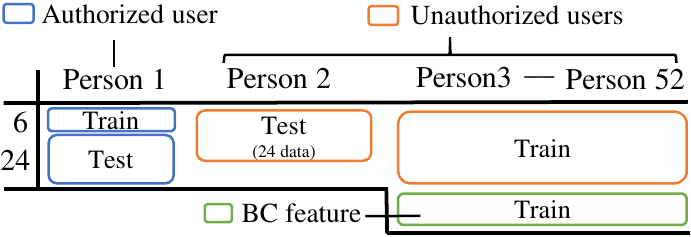}
	\caption{Training and test dataset. Person 1 is the authorized user, Person 2 is unauthorized user for test data, and Person 3 through Person 52 are the unauthorized users for learning.}
	\label{fig:dataset}
\end{figure}

In the example of Fig. \ref{fig:dataset}, Person 1 is the authorized user, Person 2 is the unauthorized user for test data, and Person 3 to Person 52 are the unauthorized users for learning.
The authorized user and the unauthorized user for test form $_{52} \mathrm{P}_2 = 2,652$ possible combinations, and the SVM learned each combination and calculated the accuracy.
For each sample of test data, the SVC function calculates the probability of being an authorized user in the range from 0 to 1.
The FAR, FRR, and TAR when the threshold is changed are calculated from the estimated probability using the roc\_curve function of scikit-learn.
To calculate the posterior probability, the probability parameter of the SVC function was set to True, and the other parameters of the SVC function were set to default. 

The ratio of BC features, $r$, was changed to 11 different values: 0.01 and 0.05, then 0.1 to 0.9 with an increment of 0.1.
The number of BC features, $N_{BC}$, was changed to 9, 45, 90, 450, 900, 4,500, and 9,000.
As an example, in Fig. \ref{fig:dataset}, when $N_{BC}$ is 9,000, there are 6 data samples of an authorized user for learning, 1,500 data samples of unauthorized users for learning, and 9,000 data samples generated by BC features for learning, totaling 10,506 data samples for learning. 
%

\section{Result}
\begin{figure}[tb]
	\centering
	\includegraphics[width=0.9\linewidth]{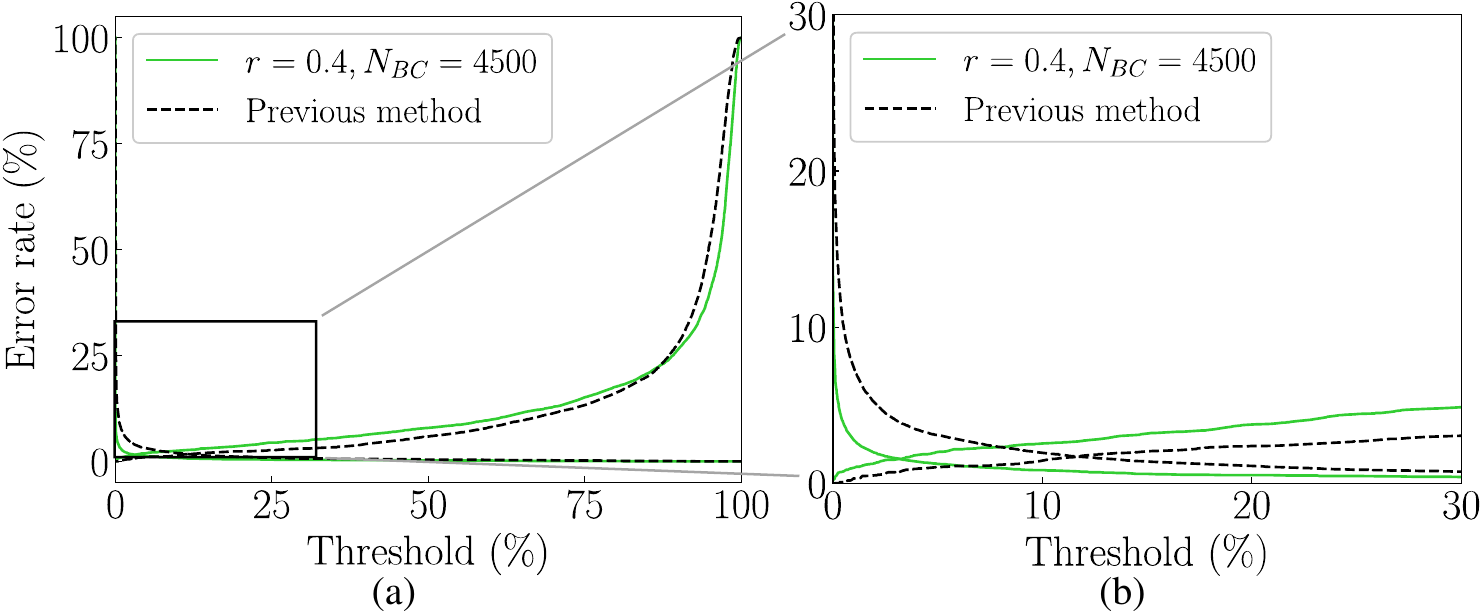}
	\caption{DET curves of $r=0.4, N_{BC}=4500$ condition. (a) overall view (b) enlarged view}
	\label{fig:det-result}
\end{figure}

Table \ref{tab:resultspk} shows the conditions that produced more accurate results than the previous method in all five evaluation indexes.
The conditions that produced the highest accuracy were $r=0.4$ and $N_{BC} = 4,500$, generating an EER of 1.56\%, which is 0.15\% lower than the previous method which produces single ear data without BC features \cite{yasuhara2019}.
 The FRR with FAR = 0.1\% was 14.35\%, or 7.95\% lower than the previous method.
Fig. \ref{fig:det-result}(a) shows the DET curve with $r=0.4$ and $N_{BC}=4,500$, and Fig. \ref{fig:det-result}(b) is an enlarged view of part of Fig. \ref{fig:det-result}(a).
As a result of learning the BC features as unauthorized users, FAR dropped sharply, and FRR increased as the threshold increased. EER moved toward lower thresholds as the FAR and FRR curves changed. 

The results with all conditions are shown in Table \ref{tab:results}.
The numbers in bold are those that were more accurate than the previous method.
The stars in the table indicate conditions that produced more accurate results than the previous method in all five evaluation indexes. 
%

\section{Discussion}
\subsection{Validity of accuracy enhancement by BC features}
Since BC features generate intermediate features of authorized and unauthorized users, they may have a negative impact on the learning of discriminators such as SVM.
Because of the trade-off between FAR and FRR, even if the introduction of BC features results in a lower FAR, it cannot be considered more accurate if EER increases or AUC decreases.
In this experiment, we introduced BC features using parameters ($r$ and $N_{BC}$) found by grid search, and as a result, four sets of conditions produced more accurate results (all evaluation indexes) than the previous method.
These results indicate that by setting appropriate BC features parameters, $r$ and $N_{BC}$, we were able to enhance the authentication accuracy by decreasing not only FAR but also EER without impacting the learning of the discriminator (SVM in this case). 
%

\subsection{Comparison between the expected DET curve and the DET curve obtained from the experiment}
A comparison between the expected DET curve, shown in Fig. \ref{fig:bc-det}, and the DET curve obtained from the experiment, in Fig. \ref{fig:det-result}, shows that in both graphs, FAR rapidly decreased and FRR increased as the threshold increased.
EER also moved toward smaller thresholds with the changes in the FAR and FRR curves.
Based on this, it can be said that the proposed method successfully reduced FAR by learning BC features as unauthorized users. 

\subsection{Changes in DET curves due to differences in BC feature parameters}
\begin{figure}[tb]
	\centering
	\includegraphics[width=0.9\linewidth]{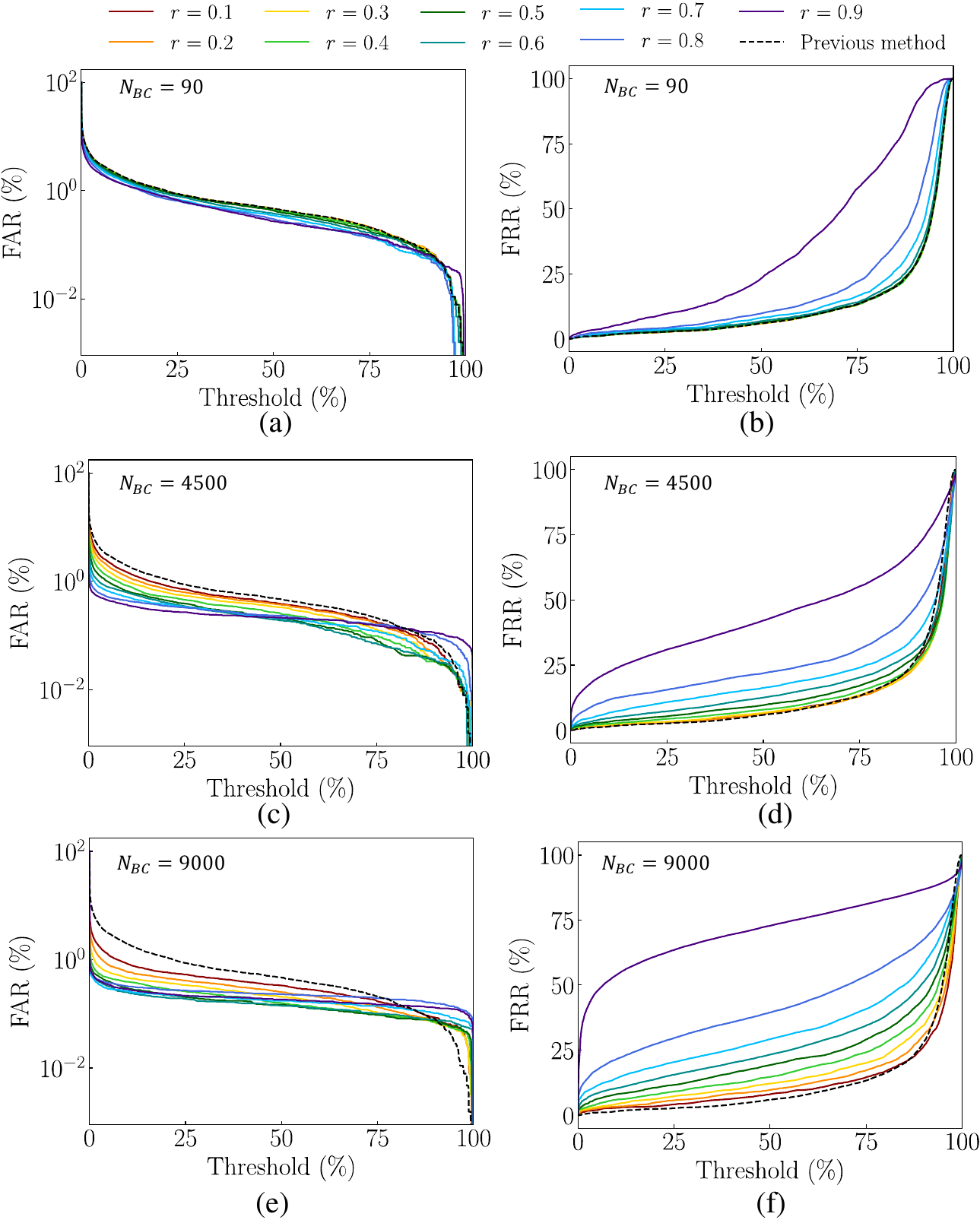}
	\caption{FAR and FRR curves with various parameters. (a) FAR curve of $N_{BC}=90$. (b) FRR curve of $N_{BC}=90$. (c) FAR curve of $N_{BC}=4500$. (d) FRR curve of $N_{BC}=4500$. (e) FAR curve of $N_{BC}=9000$. (f) FRR curve of $N_{BC}=9000$.}
	\label{fig:farfrr-result}
\end{figure}

Fig. \ref{fig:farfrr-result} shows DET curves when $N_{BC}$ is 90, 4,500, and 9,000.
The FAR curves when BC features are introduced show no significant difference between an $N_{BC}$ of 4,500 and 9,000.
On the other hand, the FRR curves show bigger differences compared to the previous method as $N_{BC}$ increases.
We expected that as $N_{BC}$ was increased, FAR would continue to decrease and FRR would continue to increase, but in practice, FAR did not continue to decrease when $N_{BC}$ was increased.
With an $N_{BC}$ of 4,500 and 9,000, FAR decreased significantly, but there was no big difference between these two conditions.
On the other hand, FRR was higher when $N_{BC}$ was 9,000 than when it was 4,500, which makes 4,500 a better parameter.
In this experiment, all four conditions that improved all accuracy evaluation indexes had an $N_{BC}$ of 4,500. Based on this, it is possible to conclude that the $N_{BC}$ of 4,500 is the ideal parameter that can reduce FAR with a minimal increase in FRR. 

As for $r$, the FAR when $N_{BC}$ is 90 and 4,500 barely changes until $r$ reaches 0.4. On the other hand, the FRR graph when $N_{BC}$ is 4,500 shows that the larger the $r$, the lower the FAR.
In this experiment, the four conditions that enhanced all accuracy evaluation indexes had an $r$ of 0.1, 0.2, 0.3, or 0.4.
Therefore, an $N_{BC}$ of 4,500 and an $r$ between 0.1 and 0.4 were the right parameters that reduced FAR without increasing FRR. 
%

\begin{table*}[tb]
	\centering
	\caption{Classification performance of conditions that produced more accurate results than the previous method in all five evaluation indexes.}
	\label{tab:resultspk}
	\begin{tabular}{ccccccc}
		\hline
		$r$ & $N_{BC}$ & AUC & EER (\%) & FRR at FAR=0.01 (\%) & FRR at FAR=0.1 (\%) & FRR at FAR=1 (\%)\\
		\hline
		\multicolumn{2}{c}{Previous method} & 0.9983  & 1.71  & 78.05  & 22.30  & 2.45 \\
		\hline
		0.1 & 4500 & {0.9984}  & {1.62}  & {69.53}  & {20.01}  & {2.36} \\
		0.2 & 4500 & {0.9985}  & {1.61}  & {56.23}  & 17.93  & {2.15} \\
		0.3 & 4500 & {0.9984}  & {1.56}  & {56.10}  & {16.04}  & {2.20} \\
		0.4 & 4500 & {0.9985}  & {1.61}  & {56.23}  & {17.93}  & {2.15} \\
		\hline
	\end{tabular}
\end{table*}
%

\section{Conclusions}
One problem reported in ear acoustic authentication, which is a type of biometric authentication, is that it yields a high FRR when FAR = 0.1\%; that is, when the security level is increased, the ear acoustic authentication becomes less convenient.
Therefore, in this study, we proposed a method to enhance the accuracy of ear acoustic authentication systems by reducing the FRR when FAR = 0.1\% based on BC features.
BC features are intermediate features of authorized and unauthorized users.
We hypothesized that if a discriminator (SVM) learned BC features as unauthorized users, the number of unauthorized users incorrectly recognized as authorized users would decrease; and with a lower FAR, the FRR when FAR = 0.1\% would also decrease.
The results were as hypothesized; the use of BC features decreased the FRR by 7.95\% when FAR = 0.1\%, with the EER being reduced by 0.15\%, and the acoustic authentication accuracy improved.
Additionally, a grid search was required to obtain BC feature parameters suitable for the dataset when generating BC features.
In this study, we devised BC features to reduce the FRR of ear acoustic authentication, and achieved decreases in the FRR and improvements in the classification accuracy.
These BC features may be applicable to other biometrics where there are trade-offs between FAR and FRR; thus, further studies need to be conducted.

\begin{table}[tb]
	\centering
	\caption{Classification performance with varying $r$ and $N_{BC}$. The star * indicates the performance is beyond previous method by all evaluation factors.}
	\label{tab:results}
	\scalebox{0.7}{ 
	\begin{tabular}{ccccccc}
		\hline
		$r$ & $N_{BC}$ & AUC & EER (\%) & FRR at FAR=0.01 (\%) & FRR at FAR=0.1 (\%) & FRR at FAR=1 (\%)\\
		\hline
		\multicolumn{2}{c}{Previous method} & 0.9983  & 1.71  & 78.05  & 22.30  & 2.45 \\
		\hline
		 & 9 & 0.9983  & \textbf{1.68}  & 78.84  & \textbf{22.10}  & 2.46 \\
		 & 45 & 0.9983  & \textbf{1.70}  & \textbf{75.93}  & \textbf{21.83}  & 2.46 \\
		 & 90 & 0.9983  & \textbf{1.67}  & \textbf{74.74}  & 22.31  & \textbf{2.40} \\
		 0.01 & 450 & 0.9983  & \textbf{1.66}  & 78.20  & 22.40  & \textbf{2.40} \\
		 & 900 & 0.9983  & \textbf{1.65}  & \textbf{76.57}  & 23.50  & \textbf{2.40} \\
		 & 4500 & 0.9983  & \textbf{1.62}  & \textbf{72.68}  & \textbf{22.10}  & 2.51 \\
		 & 9000 & 0.9983  & 1.73  & 82.35  & 23.83  & 2.74 \\
		 & 9 & 0.9983  & \textbf{1.68}  & 78.66  & \textbf{22.10}  & 2.46 \\
		 & 45 & 0.9983  & \textbf{1.70}  & \textbf{75.91}  & 22.33  & 2.46 \\
		 & 90 & 0.9983  & \textbf{1.66}  & \textbf{76.07}  & \textbf{22.28}  & \textbf{2.40} \\
		 0.05 & 450 & 0.9983  & \textbf{1.66}  & \textbf{76.16}  & \textbf{22.15}  & \textbf{2.40} \\
		 & 900 & 0.9983  & \textbf{1.65}  & \textbf{74.14}  & 23.25  & \textbf{2.40} \\
		 & 4500 & \textbf{0.9984}  & \textbf{1.61}  & \textbf{69.66}  & \textbf{21.27}  & 2.50 \\
		 & 9000 & 0.9983  & 1.73  & 82.78  & 23.00  & 2.69 \\
		 & 9 & 0.9983  & \textbf{1.68}  & 78.81  & \textbf{22.06}  & 2.46 \\
		 & 45 & 0.9983  & \textbf{1.70}  & \textbf{76.40}  & \textbf{22.20}  & 2.46 \\
		 & 90 & 0.9983  & \textbf{1.66}  & \textbf{75.12}  & \textbf{22.27}  & \textbf{2.40} \\
		 0.1 & 450 & 0.9983  & \textbf{1.66}  & \textbf{73.16}  & \textbf{22.07}  & 2.46 \\
		 & 900 & 0.9983  & \textbf{1.65}  & \textbf{71.09}  & 22.71  & \textbf{2.39} \\
		 & *4500 & \textbf{0.9984}  & \textbf{1.62}  & \textbf{69.53}  & \textbf{20.01}  & \textbf{2.36} \\
		 & 9000 & 0.9983  & 1.72  & 85.19  & \textbf{21.90}  & 2.65 \\
		 & 9 & 0.9983  & \textbf{1.68}  & 78.63  & \textbf{22.07}  & 2.46 \\
		 & 45 & 0.9983  & \textbf{1.70} & \textbf{76.17}  & 22.60  & 2.46 \\
		 & 90 & 0.9983  & \textbf{1.66}  & \textbf{76.40}  & \textbf{22.21}  & \textbf{2.40} \\
		 0.2 & 450 & 0.9982  & \textbf{1.64}  & \textbf{70.48}  & \textbf{21.67}  & 2.47 \\
		 & 900 & 0.9983  & \textbf{1.66}  & \textbf{69.17}  & \textbf{21.43}  & \textbf{2.41} \\
		 & *4500 & \textbf{0.9985}  & \textbf{1.61}  & \textbf{56.23}  & \textbf{17.93}  & \textbf{2.15} \\
		 & 9000 & \textbf{0.9984}  & 1.71  & 88.47  & \textbf{20.88}  & \textbf{2.22} \\
		 & 9 & 0.9983  & \textbf{1.69}  & 78.98  & \textbf{22.13}  & 2.46 \\
		 & 45 & 0.9983  & \textbf{1.70} & 78.53  & 22.48  & 2.46 \\
		 & 90 & 0.9983  & 1.71  & \textbf{77.47}  & \textbf{21.62}  & \textbf{2.40} \\
		 0.3 & 450 & 0.9981  & \textbf{1.68}  & \textbf{72.64}  & \textbf{19.47}  & \textbf{2.40} \\
		 & 900 & 0.9982  & \textbf{1.69}  & \textbf{68.60}  & \textbf{18.87}  & \textbf{2.40} \\
		 & *4500 & \textbf{0.9984}  & \textbf{1.56}  & \textbf{56.10}  & \textbf{16.04}  & \textbf{2.20} \\
		 & 9000 & 0.9983  & \textbf{1.67}  & 90.59  & \textbf{21.12}  & \textbf{2.12} \\
		 & 9 & 0.9983  & \textbf{1.70} & 79.56  & \textbf{21.90}  & 2.46 \\
		 & 45 & 0.9983  & 1.73  & 79.38  & \textbf{22.01}  & 2.48 \\
		 & 90 & \textbf{0.9984}  & 1.72  & 78.23  & \textbf{21.10}  & 2.53 \\
		 0.4 & 450 & 0.9980  & \textbf{1.68}  & \textbf{66.61}  & \textbf{19.16}  & \textbf{2.42} \\
		 & 900 & 0.9981  & \textbf{1.68}  & \textbf{67.38}  & \textbf{19.31}  & \textbf{2.40} \\
		 & *4500 & \textbf{0.9984}  & \textbf{1.56}  & \textbf{56.68}  & \textbf{14.35}  & \textbf{2.29} \\
		 & 9000 & 0.9983  & \textbf{1.59}  & 95.73  & 23.92  & \textbf{2.12} \\
		 & 9 & 0.9983  & \textbf{1.68}  & 80.51  & \textbf{21.85}  & \textbf{2.41} \\
		 & 45 & 0.9983  & \textbf{1.67}  & 79.49  & \textbf{21.88}  & 2.55 \\
		 & 90 & \textbf{0.9984}  & \textbf{1.70} & 78.14  & \textbf{21.36}  & 2.57 \\
		 0.5 & 450 & 0.9980  & \textbf{1.67}  & \textbf{64.43}  & \textbf{19.68}  & 2.48 \\
		 & 900 & 0.9981  & \textbf{1.66}  & \textbf{64.55}  & \textbf{19.62}  & \textbf{2.39} \\
		 & 4500 & 0.9982  & \textbf{1.70} & \textbf{60.65}  & \textbf{15.42}  & 2.49 \\
		 & 9000 & 0.9981  & \textbf{1.51}  & 96.83  & 27.18  & \textbf{2.05} \\
		 & 9 & \textbf{0.9984}  & \textbf{1.69}  & \textbf{71.40}  & \textbf{21.40}  & 2.63 \\
		 & 45 & 0.9983  & 1.81  & \textbf{73.32}  & 23.07  & 2.80 \\
		 & 90 & 0.9983  & 1.81  & \textbf{74.38}  & \textbf{20.77}  & 2.88 \\
		 0.6 & 450 & 0.9978  & 1.76  & \textbf{71.89}  & \textbf{20.10}  & 2.78 \\
		 & 900 & 0.9980  & 1.73  & \textbf{74.38}  & \textbf{20.44}  & 2.75 \\
		 & 4500 & 0.9979  & \textbf{1.70} & \textbf{68.22}  & \textbf{17.43}  & 2.65 \\
		 & 9000 & 0.9979  & \textbf{1.44}  & 99.52  & 35.83  & \textbf{1.89} \\
		 & 9 & 0.9983  & 1.71  & \textbf{76.59}  & \textbf{21.96}  & 2.64 \\
		 & 45 & 0.9982  & 1.92  & 82.12  & \textbf{21.15}  & 3.23 \\
		 & 90 & 0.9981  & 1.93  & 81.78  & \textbf{19.76}  & 3.25 \\
		 0.7 & 450 & 0.9976  & 1.90  & \textbf{64.34}  & 24.26  & 3.20 \\
		 & 900 & 0.9978  & 1.83  & \textbf{64.44}  & \textbf{21.12}  & 3.17 \\
		 & 4500 & 0.9975  & \textbf{1.59}  & 81.07  & 26.51  & 2.98 \\
		 & 9000 & 0.9975  & \textbf{1.46}  & 99.91  & 60.60  & \textbf{1.91} \\
		 & 9 & 0.9980  & 1.99  & 82.78  & 28.83  & 2.97 \\
		 & 45 & 0.9979  & 2.05  & 92.40  & 26.80  & 3.42 \\
		 & 90 & 0.9977  & 2.14  & 89.38  & 29.47  & 3.29 \\
		 0.8 & 450 & 0.9972  & 2.12  & 85.45  & 28.01  & 3.59 \\
		 & 900 & 0.9972  & 1.95  & 85.08  & 29.52  & 3.41 \\
		 & 4500 & 0.9968  & 1.85  & 92.72  & 49.41  & 3.46 \\
		 & 9000 & 0.9963  & 1.75  & 99.92  & 89.58  & 3.91 \\
		 & 9 & 0.9959  & 2.86  & 94.62  & 50.48  & 6.77 \\
		 & 45 & 0.9961  & 2.45  & 99.78  & 59.90  & 6.44 \\
		 & 90 & 0.9960  & 2.53  & 100.00  & 69.56  & 6.53 \\
		 0.9 & 450 & 0.9955  & 2.43  & 100.00  & 69.90  & 6.11 \\
		 & 900 & 0.9952  & 2.17  & 100.00  & 94.92  & 5.65 \\
		 & 4500 & 0.9953  & 2.05  & 99.34  & 76.87  & 6.70 \\
		 & 9000 & 0.9887  & 4.70  & 99.77  & 92.23  & 11.87 \\
		 \hline
	\end{tabular}}
\end{table}

\section*{Acknowledgments}
This work was supported by JSPS KAKENHI Grant Numbers JP19H04112, JP19K22851.

\bibliographystyle{unsrtnat}
\bibliography{yasuhara}  






\end{document}